\documentclass[twocolumn,showpacs,preprintnumbers,amsmath,amssymb]{revtex4}
\usepackage{graphicx}
\usepackage{dcolumn}
\usepackage{bm}
\newcommand{\be}{\begin{equation}}
\newcommand{\ee}{\end{equation}}
\newcommand{\bea}{\begin{eqnarray}}
\newcommand{\eea}{\end{eqnarray}}

\begin{document}

\title{Ground state and finite temperature signatures of quantum
phase transitions\\
in the half-filled Hubbard model on a honeycomb lattice}

\author{
Thereza Paiva,$^1$ R.T.~Scalettar,$^2$ W.~Zheng,$^3$, R.R.P.~Singh,$^2$
and J. Oitmaa$^3$
}
\affiliation{$^1$ Instituto de F\' \i sica,
      Universidade Federal do Rio de Janeiro,
                Cx.P.\ 68.528,
      21945-970 Rio de Janeiro RJ, Brazil}
\affiliation{$^2$ Physics Department, University of California, Davis, CA 95616}
\affiliation{$^3$ School of Physics, Univ.~of New South Wales, Sydney NSW 2052,
Australia}

\date{\today}

\begin{abstract}
We investigate ground state and finite temperature properties
of the half-filled Hubbard
model on a honeycomb lattice using quantum monte carlo
and series expansion techniques.  Unlike  the square lattice, for which
magnetic order exists at $T=0$ for any non-zero $U$, the honeycomb lattice
is known to have a
semi-metal phase at small $U$ and an antiferromagnetic one at large $U$.
We investigate the phase transition at $T=0$ by studying the magnetic structure factor
and compressibility using quantum monte carlo simulations and by
calculating the sublattice magnetization, uniform susceptibility,
spin-wave and single hole 
dispersion using series
expansions around the ordered phase.
Our results are consistent with a single continuous
transition at $U_c/t$ in the range $4-5$.
Finite temperature signatures of this phase transition
are seen in the behavior of the specific heat, $C(T)$, which changes from
a two-peaked structure for $U>U_c$ to a one-peaked structure for
$U < U_c$.
Furthermore, the $U$ dependence of the
low temperature coefficient of $C(T)$ exhibits an anomaly
at $U \approx U_c$.
\end{abstract}
\pacs{71.10.Fd,75.10.Lp,75.40.Mg}
\maketitle

\section{ Introduction}

The two-dimensional Hubbard Hamiltonian has been
extensively studied as a model of metal--insulator and magnetic
phase transitions\cite{Montorsi91} and also within the context of
systems such as the CuO$_2$ sheets of high temperature
superconductors.\cite{Scalapino94} In the square-lattice case,
at half-filling,
nesting of the Fermi surface leads to a divergent antiferromagnetic
susceptibility as the temperature is lowered, even for $U=0$,
and thus the ground state is an antiferromagnetic
insulator at any non-zero $U$.
It is of interest to study cases where, instead, the transition
to the antiferromagnetic phase occurs at finite $U$.
In such a situation, for example, it may
prove possible to see if the Mott metal-insulator and
paramagnetic--antiferromagnetic phase transitions occur separately.
A finite $U_c$ also makes it more straightforward to
study the thermodynamics at temperatures above the
$T=0$ quantum phase transition, a question pertinent to experimental
studies of such phase transitions.

The two--dimensional honeycomb lattice is one
geometry in which we can explore these issues.
In this paper, we investigate the properties of the
half--filled honeycomb lattice
Hubbard model using determinant
quantum monte carlo and series expansions methods.
After a brief review of previous work,
we describe the model and calculational approaches,
and show data for a number of different ground state
properties that carry signatures of the phase transition.
Our overall results are consistent with a single continuous
transition as a function of $U/t$.
We then turn to the finite temperature behavior of
the specific heat to see how such a critical point may be reflected
in this key experimental property.

While the honeycomb lattice has $U_c$ non-zero, it is important
to note at the outset that,
like the square lattice, its
non-interacting density of states has a special feature.
As shown in Fig.~1,
$N(\omega)$ vanishes linearly
as $\omega \rightarrow 0$, so the system is a
semi--metal (or alternatively,
a zero-gap semiconductor) at half--filling.
As a consequence,  at weak coupling, the low temperature behavior of the
specific heat is quadratic in temperature, $C = \delta T^2$, instead of the
usual linear Fermi liquid dependence.
At strong coupling, when long range antiferromagnetic order
sets in, the specific heat will also be quadratic in $T$
owing to the spin--wave excitations.
How the specific heat evolves between these two regimes is an open question.

A considerable body of work exists concerning the ground state
phase diagram.  Martelo {\it et al}
found that within mean field theory
the Mott transition occurs below an upper bound for
the critical interaction
strength $U_{\rm c}/t \approx 5.3$.\cite{Martelo97}
Meanwhile, their variational
monte carlo calculation suggested a lower bound for
the antiferomagnetic
transition $U_{\rm c}/t \approx 3.7$.
They interpreted these results as a single transition
from paramagnetic metal to antiferromagnetic insulator at
$U_{\rm c}/t = 4.5 \pm 0.5$.

Baskaran {\it et al} \cite{Baskaran02} and Sorella {\it et al} \cite{Sorella92}
studied the model using the Random Phase Approximation
which gives $U_{\rm c}/t = 2.23$ for the onset of antiferromagnetic order.
Associated auxiliary field quantum monte carlo (QMC) simulations\cite{Sorella92}
using the ground state projection approach
suggested $U_{\rm c}/t = 4.5 \pm 0.5$.
Later QMC work by Furukawa\cite{Furukawa01} on larger lattices
and doing system size extrapolations resulted in a somewhat
lower value,
$U_{\rm c}/t = 3.6 \pm 0.1$.
Peres {\it et al} have recently studied the phase diagram
and mean field magnetization of coupled honeycomb layers
as a function of filling, $U/t$, and interlayer
hopping $t'/t$ using the random phase approximation and
spin wave analysis.\cite{Peres04}

As with the square lattice Hubbard model, Nagaoka
ferromagnetism is also possible for the doped honeycomb lattice
at strong couplings ($U/t > 12$), as has been investigated by Hirsch
using exact diagonalization on small clusters\cite{Hirsch94} and by
Hanisch {\it et al} using Gutzwiller wave functions.\cite{Hanisch95}

We conclude this introduction by noting that
the Hubbard model on a honeycomb lattice has also been suggested
to be of interest for
a variety of systems including graphite sheets,\cite{Baskaran02}
and carbon nanotubes\cite{Balents97}, as well as
MgB$_2$\cite{Furukawa01} and
Pb and Sn on Ge(111) surfaces.\cite{Hellberg04}
The honeycomb lattice is also a 2/3 subset of the triangular lattice,
and so the nature of spin correlations on the honeycomb lattice
has been considered as possibly relevant to the properties of
triangular lattice systems like
Na$_x$CoO$_2$ at appropriate dopings \cite{weihong}.

\begin{figure}
\label{dos}
\includegraphics[width=2.5in,height=2.5in,angle=270]{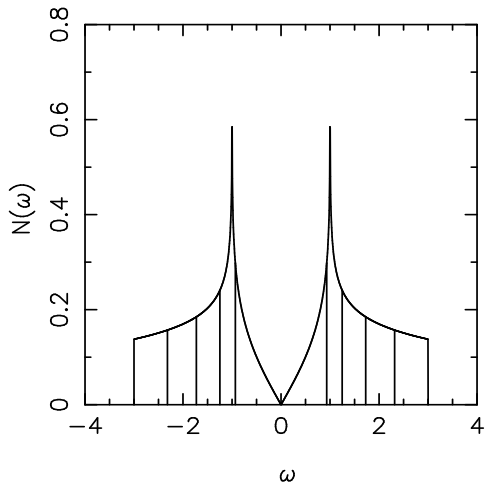}
\caption{
The noninteracting density of states of the Hubbard model
on a honeycomb lattice.
This geometry is bipartite, so
$N(\omega)$ is symmetric about $\omega=0$.
The vertical lines correspond to fillings of 0.1, 0.2, 0.3, $.\,.\,.$
The density of states vanishes linearly at $\omega=0$
and has logarithmic Van Hove singularities at fillings
$\rho=3/8$ and $5/8$.
The bandwidth $W=6t$.
}
\end{figure}

\vskip0.1in
\section{The Hubbard Hamiltonian,
Determinant Quantum Monte Carlo,
and Series Expansion Methods}

We study the Hubbard Hamiltonian,
\begin{eqnarray}
H &=& -t \sum_{\langle {\bf ij} \rangle \sigma}
(c_{{\bf i}\sigma}^{\dagger} c_{{\bf j}\sigma}^{\phantom{\dagger}}
+ c_{{\bf j}\sigma}^{\dagger} c_{{\bf i}\sigma}^{\phantom{\dagger}})
\nonumber
\\
&+& U \sum_{{\bf i}} (n_{{\bf i}\uparrow} - \frac12)
(n_{{\bf i}\downarrow} - \frac12)
- \mu \sum_{{\bf i}} (n_{{\bf i}\uparrow} + n_{{\bf i}\downarrow} ).
\nonumber
\label{hubham}
\end{eqnarray}
Here $c_{{\bf i}\sigma}^{\dagger} (c_{{\bf j}\sigma}^{\phantom{\dagger}})$
are creation(destruction) operators for
a fermion of spin $\sigma$ on lattice site ${\bf i}$.
The kinetic energy term includes a sum
over near neighbors $\langle {\bf i},{\bf j} \rangle$ on a
two--dimensional honeycomb lattice.  We denote by $N$
the number of lattice sites, and $L$ the linear dimension with
$N = \frac23 L^2$.  The
interaction term is written in particle--hole
symmetric form so that $\mu=0$ corresponds to
half--filling:  the density
$\rho=\langle n_{ {\bf i}\uparrow}+n_{ {\bf i}\downarrow}
\rangle=1$
for all $t,U$ and temperatures $T$.
We choose the hopping parameter $t=1$ to set the energy scale.
Note that the noninteracting model has
two `Dirac' points $\vec K_\pm$ on the Fermi surface
where the dispersion relation is relativistic,\cite{Baskaran02} i.e. the
energy grows linearly with $|\vec k - \vec K_\pm|$.

We use the determinant quantum monte carlo method to
measure the properties of the Hamiltonian.\cite{Blankenbecler81}
In this approach the partition function is
written as a path integral, the interaction term is decoupled through
the introduction of a discrete auxiliary `Hubbard-Stratonovich'
field,\cite{Hirsch85}
and the fermion degrees of freedom are traced out analytically.
The remaining summation over the Hubbard-Stratonovich field
is done stochastically.
Since the lattice is bipartite, no sign problem occurs
at half-filling.  Data were typically generated by doing
several tens of thousands of measurements at each
data point (temperature, coupling constant, lattice size).
`Global moves' which flip the Hubbard-Stratonovich
variables for all imaginary times at a given spatial site
were included so that at stronger couplings, transitions between
different densities are facilitated.\cite{Noack91}

We have also carried out an Ising type expansion for this system
at $T=0$ using a linked-cluster method.\cite{gel00}
Similar expansions were previously done
for the Hubbard model on the square lattice.\cite{rajiv_sq}
To perform the series expansion, one needs to introduce an Ising
interaction into the Hubbard Hamiltonian,
and divide the Hamiltonian into an unperturbed
 Hamiltonian ($H_0$)
and a perturbation ($H_1$) as follows,
\bea
H &=& H_0 + \lambda H_1  \nonumber \\
H_0 &=& J \sum_{\langle {\bf ij}\rangle}  ( \sigma_{\bf i}^z \sigma_{\bf j}^z  + 1)
+ U \sum_{{\bf i}} (n_{{\bf i}\uparrow} - \frac12)
(n_{{\bf i}\downarrow} - \frac12)  \nonumber  \\
H_1 &=& \sum_{\langle {\bf ij}\rangle} [ - J ( \sigma_{\bf i}^z \sigma_{\bf j}^z  + 1)
    - t  ( c_{{\bf i}\sigma}^\dag c_{{\bf j}\sigma} + {\rm h.c.} ) ]
\nonumber
\eea
where $\sigma_{\bf i}^z = n_{{\bf i}\uparrow} - n_{{\bf i}\downarrow}$,
and $\lambda$ is the expansion parameter. Note that we are primarily
interested in the behavior of the system at $\lambda=1$, at which point
the Ising term cancels between $H_0$ and $H_1$. The strength of the Ising
interaction can be varied to improve convergence, and it proves useful
to keep it of order $t^2/U$.\cite{rajiv_sq}
The limits $\lambda=0$ and $\lambda=1$
correspond to the Ising model and the original
model, respectively. The  unperturbed ground state is
the usual Ne\'el state.
The Ising series have been calculated
to order $\lambda^{15}$ for the ground state energy,
 the staggered magnetization $M$, and the square of the
local moment $L_m$\cite{rajiv_sq}, and to order
$\lambda^{13}$ for the uniform magnetic susceptibility
$\chi_{\perp}$.
The resulting power series in $\lambda$
for $t/U=0.15$ and $J/U=0.0225$ are presented in Table I.

In addition to the ground state properties,
we also compute the spin-wave dispersion $\Delta$
(to order $\lambda^{13}$) and the dispersion $\Delta_{1h}$
of 1-hole
doped to half-filled system (to order $\lambda^{11}$).
The calculation of the dispersion involves a list of 28811
clusters up to 13 sites.
The series for the dispersions can be written in the following form
\bea
\Delta (k_x,k_y)& =& \sum_p \sum_{i,j} \frac{1}{3} a_{i,j,p} \lambda^p
{\big \{} \cos (i k_x/2) \cos (\sqrt{3} j k_y/2)  \nonumber \\
 & +& \cos [ (i + 3j) k_x/4] \cos[ \sqrt{3} (i - j) k_y)/4] \nonumber \\
 & +& \cos[(i - 3j) k_x/4] \cos[\sqrt{3} (i + j)k_y/4] {\big \} }
\nonumber
\eea
In Table II, we list the
series coefficients $a_{i,j,p}$ for $t/U=0.15$ and $J/U=0.0225$.
The series for other couplings  are available from the authors
upon request.

\section{Quantum Phase Transition}

We begin by examining the evidence for a phase transition
in the model. First, we present results from the quantum monte carlo
simulations, which can, in principle, address arbitrary $t/U$ ratios.

\subsection{Compressibility}

The Mott metal--insulator transition is signalled by a vanishing
compressibility $\kappa=\partial \rho / \partial \mu$.
We show $\rho$ as a function of $\mu$ for $\beta=8$ and $U=2t$ (Fig.~2) and
$U=7t$ (Fig.~3).  There is a clear qualitative difference in behavior.
For weak coupling, $\rho$ immediately shifts from half-filling as $\mu$
increases from zero, while at strong coupling, $\rho$ remains pinned at
$\rho=1$ out to finite $\mu$.

In Fig.~4 we show the filling as a function of $U$
at a small non-zero value of the chemical potential
$\mu_0=0.20$.  We see that $\rho_0 \rightarrow 1$ at $U \approx 5t$,
signalling the onset of the Mott insulating phase.

\begin{figure}
\label{nvsmuU2}
\includegraphics[width=2.5in,height=2.5in,angle=270]{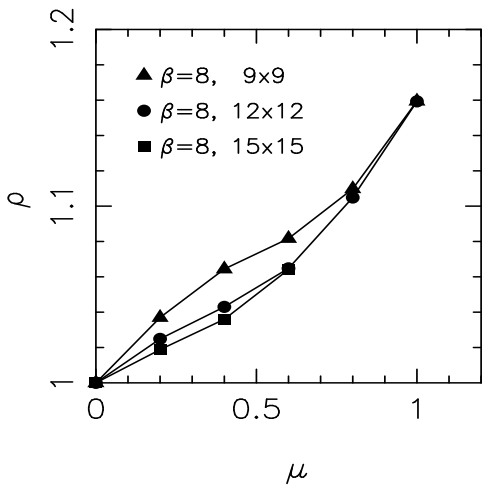}
\caption{
Density $\rho$ as a function of chemical potential $\mu$
at weak coupling ($U=2t$).  $\rho$ is not pinned at one, but immediately
begins to shift when $\mu \neq 0$:  there is no Mott gap.
}
\end{figure}

\begin{figure}
\label{nvsmuU7}
\includegraphics[width=2.5in,height=2.5in,angle=270]{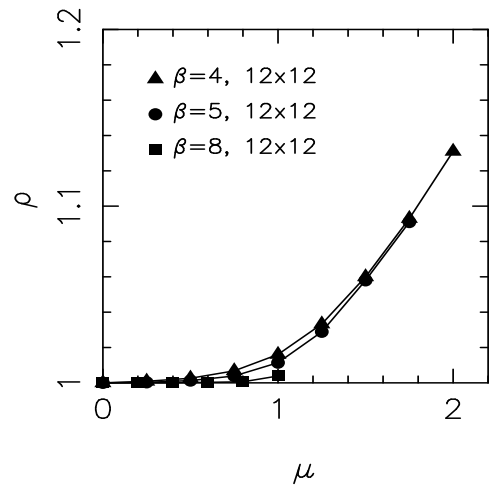}
\caption{
Density $\rho$ as a function of chemical potential $\mu$
at strong coupling ($U=7t$).  There is clear evidence for
a Mott gap.
}
\end{figure}

\begin{figure}
\label{rhovsU}
\includegraphics[width=2.5in,height=2.5in,angle=270]{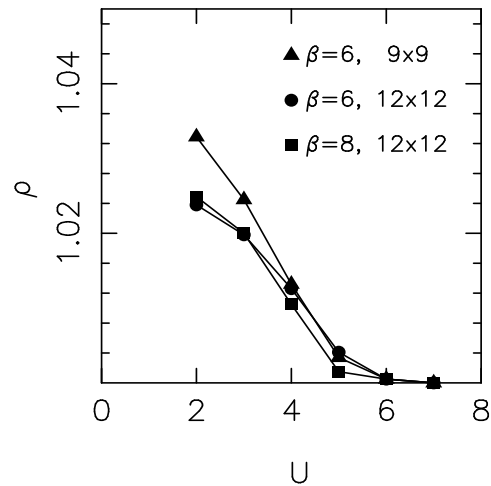}
\caption{
The difference in the value of the density $\rho$ from half-filling
at small nonzero $\mu_0=0.2t$ is a measure of the presence
the Mott gap.  Here we see $\rho(\mu_0) \rightarrow 1$
and hence a gap opens at $U/t \approx 5$.
}
\end{figure}

\subsection{Spin Correlations and Antiferromagnetic Susceptibility}

To study the magnetic behavior,
we measure the real space spin correlations,
\begin{eqnarray}
c_{zz} ({\bf r}) &=& \langle S_z({\bf r}) S_z({\bf 0}) \rangle
\hskip0.50in S_z({\bf r}) = n_{{\bf r}\uparrow} - n_{{\bf r}\downarrow}
\nonumber \\
c_{+-} ({\bf r}) &=& \langle S_-({\bf r}) S_+({\bf 0}) \rangle
\hskip0.45in S_+({\bf r}) = c_{{\bf r}\uparrow}^{\dagger} c_{{\bf r}\downarrow},
\nonumber
\end{eqnarray}
and their Fourier transforms,
\begin{eqnarray}
S_{zz}({\bf q}) = \sum_{\bf r} e^{i {\bf q}\cdot {\bf r} }
c_{zz} ({\bf r})
\nonumber \\
S_{+-}({\bf q}) = \sum_{\bf r} e^{i {\bf q}\cdot {\bf r} }
c_{+-} ({\bf r}).
\nonumber
\end{eqnarray}

At $T=0$ and in the antiferromagnetically ordered phase
at large $U/t$, the real space
correlation will go asymptotically to a non-zero value $m^2/3$
at large separations
${\bf r}$.
In our finite temperature simulations, we access the $T=0$ limit
by cooling the system to the point where the correlation length
exceeds the lattice size.
In this case, the structure factor will grow linearly with lattice size $N$.
More precisely, the structure factor will obey,
\begin{eqnarray}
{1 \over N} S({\bf q)} = m^2/3+ a/L,
\nonumber
\end{eqnarray}
where $L$ is the linear lattice size.\cite{Huse88}
In the paramagnetic phase at small $U/t$,
the structure factor will be independent of $N$, and
hence $S({\bf q})/N$ will vanish as $N \rightarrow \infty$.

In Fig.~5 we show
$S(\pi,\pi)$ as a function of inverse temperature $\beta$
for different lattice sizes at $U/t=7$.
Fig.~6 shows the associated scaling plot.
Also shown in Fig.~6 is the value of $c({\bf r})$ for the
largest separations on our finite lattices.  This quantity
should scale
with the same intercept $m^2/3$ but a different finite size correction.
We see that the system is in an antiferromagnetically ordered
phase for this coupling.

Figs.~7-8 show analogous plots at $U/t=6$.
The order parameter is still non-zero, but is quite small.
Similar plots for $U/t=5$ are consistent with the vanishing
of long range order.
While we cannot pin down the location of the
quantum phase transition exactly, a comparison of this analysis
with the compressibility of Fig.~3 suggests that the vanishing of
the compressibility gap and the antiferromagnetic order occur
very close to each other and are in the neighborhood of $U_c \approx 5t$.

\begin{figure}
\label{s7}
\includegraphics[width=2.5in,height=2.5in,angle=270]{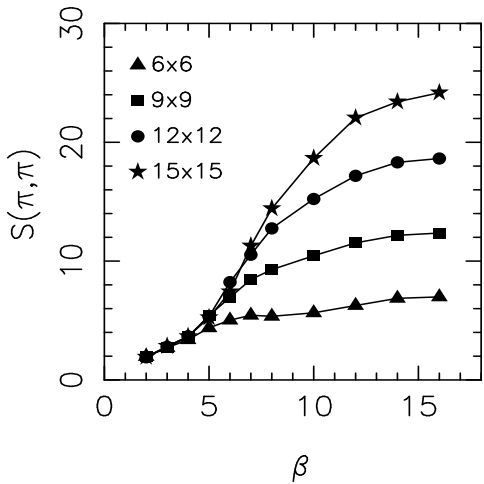}
\caption{
The antiferromagnetic structure factor is shown as a function
of inverse temperature $\beta$ and different lattices sizes $L$ at $U/t=7$.
At low $\beta$ (high $T$) the correlation length is short,
and $S(\pi,\pi)$ is independent of $L$.  At large $\beta$,
$S(\pi,\pi)$ grows with $L$.
}
\end{figure}

\begin{figure}
\label{s7scaled}
\includegraphics[width=2.5in,height=2.5in,angle=270]{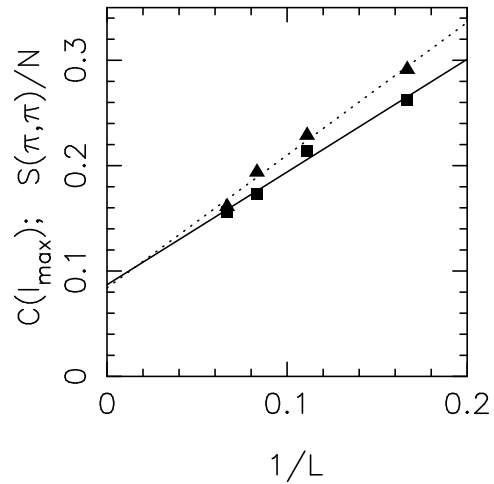}
\caption{
The scaled structure factor (filled triangles)
and spin correlation function (filled squares) at large distance
are shown for large $\beta$ as a function of the inverse linear dimension
for $U/t=7$.  The lines are least squares
fits to the data.  These quantities scale to a nonzero value of the
order parameter (square of the staggered magnetization)
in the thermodynamic limit  $1/L \rightarrow 0$.}
\end{figure}

\begin{figure}
\label{s6}
\includegraphics[width=2.5in,height=2.5in,angle=270]{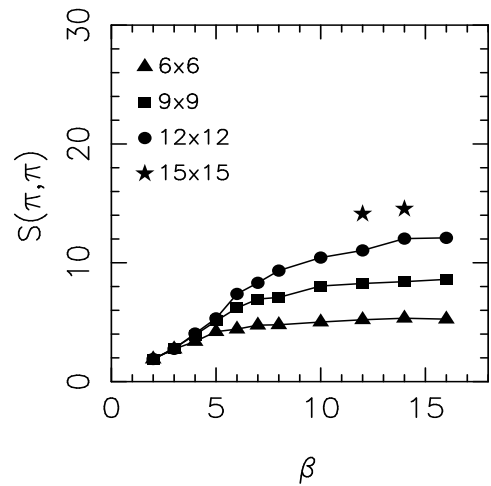}
\caption{
Same as Fig.~5 except $U/t=6$.
}
\end{figure}

\begin{figure}
\label{s6scaled}
\includegraphics[width=2.5in,height=2.5in,angle=270]{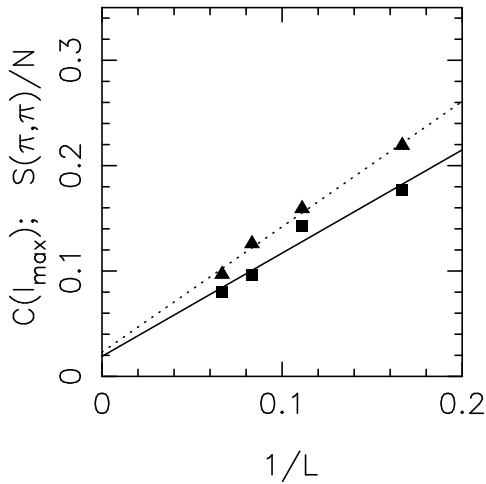}
\caption{
Same as Fig.~6 except $U/t=6$.
}
\end{figure}

\subsection{Results from Series Expansions}

We now present results from the Ising type series expansions. These
expansions are only valid in the magnetically ordered phase, and thus
can only access the properties of the system for $U>U_c$.

\begin{figure}
\includegraphics[width=8.3cm]{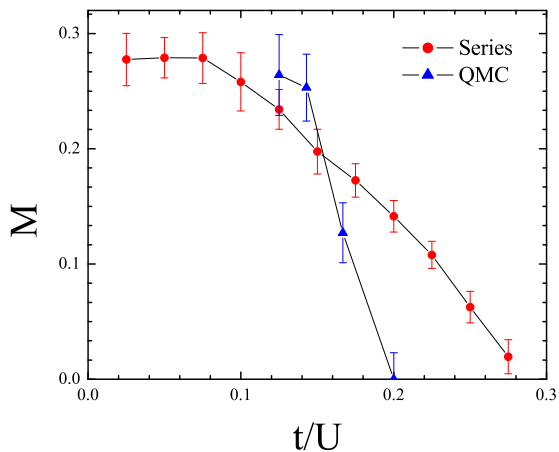}
\caption{
The staggered magnetization versus $t/U$ obtained from series expansions
and quantum monte carlo simulations.
The lines joining the points are a guide to the eye. See text
for more discussion.
}
\label{fig:M}
\end{figure}

\begin{figure}
\includegraphics[width=8.3cm]{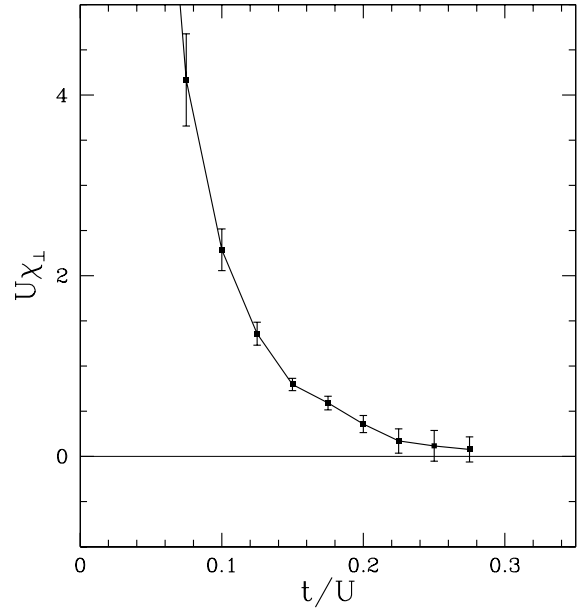}
\caption{
The uniform susceptibility $U\chi_{\perp}$ versus $t/U$ obtained from series expansion.
}
\label{fig:chi}
\end{figure}

In Fig.~\ref{fig:M} and Fig.~\ref{fig:chi},
we show the sublattice magnetization
and uniform susceptibility. The QMC results for the sublattice
magnetization are also shown. The two agree with each other for
small $t/U$. The uncertainties increase as the transition
is approached. QMC results suggest a more abrupt drop to zero around
$U/t\approx 5$, whereas the series results suggest a gradual
decrease with increasing $t/U$. Since the series are
not directly in the variable $t/U$ but rather in an auxiliary variable
$\lambda$, it is difficult to locate the true critical point $U_c/t$  and
obtain the critical properties. However, since we expect the
critical exponent $\beta$  to be less than one, the true curve
should come to zero with an infinite slope. Thus, from the series
results alone, one would estimate $U_c/t\approx 4$, and this is in
agreement with the estimate from the susceptibility $\chi_{\perp}$
shown in Fig. \ref{fig:chi}.

\begin{figure}
\includegraphics[width=8.3cm]{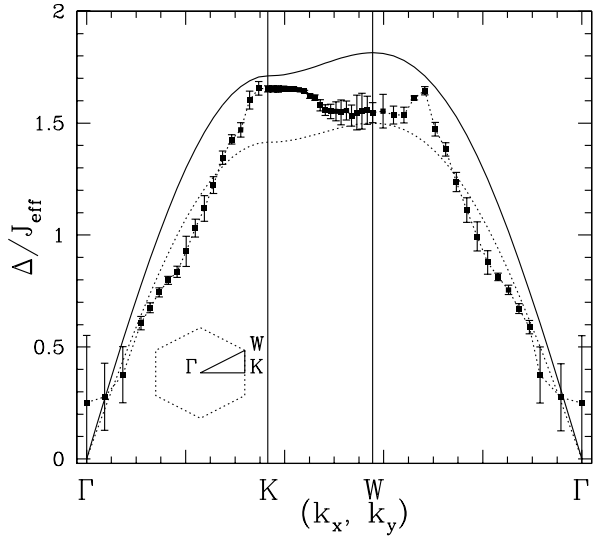}
\caption{
Plot of the spin-wave excitation spectrum
$\Delta (k_x, k_y)$ (in units of effective $J_{\rm eff}=4t^2/U$)
along the path $\Gamma K W\Gamma$ in the Brillouin zone (see the inset,
where the momentum {\bf k} for $\Gamma$, $K$, $W$
points are  $(0,0)$, $(2\pi/3,0)$, and $(2\pi/3, 2\pi/3\sqrt{3} )$,
respectively)
 for the system with coupling ratios
$t/U=0.1$ in the Ne\'el  ordered phase. Also shown are the first (dashed line) and second (solid line)
order spin-wave results\cite{sw_haf} for Heisenberg model on honeycomb lattice.
}
\label{fig:mk_spin}
\end{figure}

Next, in Fig.~\ref{fig:mk_spin}, we show the
spin-wave dispersion along  high-symmetry cuts through the Brillouin
zone for $t/U=0.1$, together with the dispersion obtained from
first and  second order spin-wave results
for the Heisenberg model on a honeycomb lattice\cite{sw_haf}, which should
approach the dispersion for the Hubbard model in the large $U$ limit. We
can see that the dispersion has its minimum
at the $\Gamma$ point.
The spin-wave dispersion for the Heisenberg model on a honeycomb
lattice has a maximum at $W$ point,
while for the Hubbard model, this is only true for very small $t/U$.
Already for $t/U=0.1$, the energy at $W$ points is reduced, and  the maximum
moves to the $K$ point.

\begin{figure}
\includegraphics[width=8.3cm]{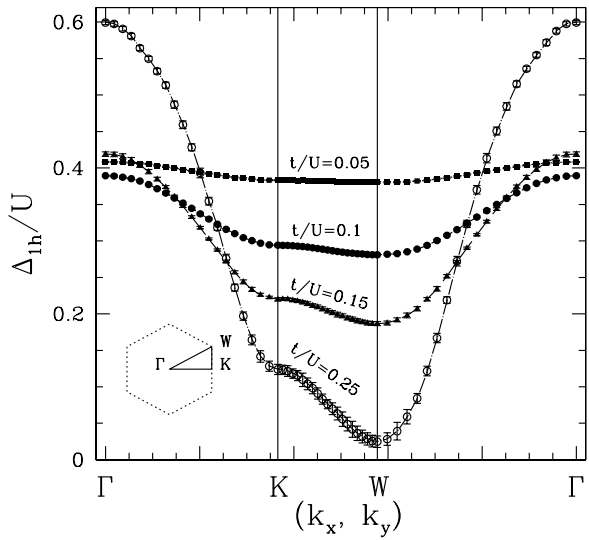}
\caption{
Plot of the  1-hole excitation spectrum
$\Delta (k_x, k_y)/U$
in the Ne\'el  ordered phase
along the path $\Gamma K W\Gamma$ in the Brillouin zone (see the inset)
 for the system with coupling ratios
$t/U=0.05$, 0.01, 0.15, 0.25.
}
\label{fig:mk_1h}
\end{figure}

\begin{figure}
\includegraphics[width=8.3cm]{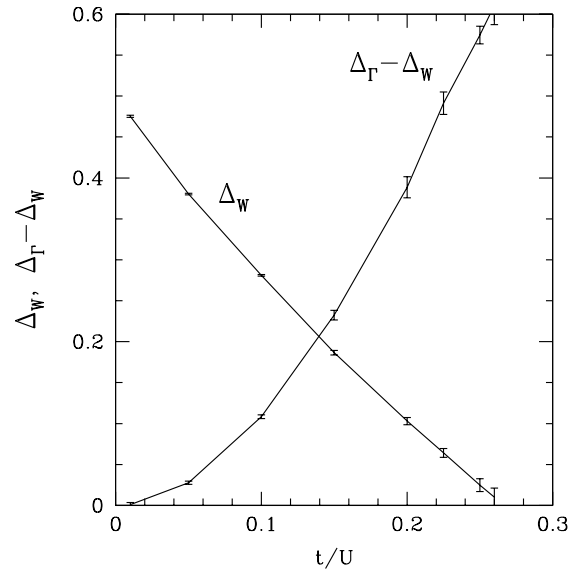}
\caption{
The minimum single-hole gap $\Delta_W$ at $W$ point and
its bandwidth $\Delta_{\Gamma}-\Delta_{W}$
 vs $t/U$.
}
\label{fig:gap_bandwidth}
\end{figure}

Also, in figure \ref{fig:mk_1h} we show the 1-hole dispersion for selected values
of $t/U$, where we can see that
the minimum and maximum gaps are at the $W$ and $\Gamma$ points, respectively.
This dispersion is quite different from the case of the square
lattice, since there is no nesting of the Fermi surface here.
For the square lattice,
the single hole dispersion relation is anomalously flat near
the degenerate points  $(0,\pm \pi)$, $(\pm \pi,0)$ of the Brillouin zone,
with the minimum of the dispersion at $(\pm \pi/2, \pm \pi/2)$.\cite{sq_1h}
Fig. \ref{fig:gap_bandwidth} shows the minimum gap,
i.e.~the gap at the $W$ point, and
the bandwidth, $\Delta_{\Gamma}-\Delta_{W}$, vs $t/U$. The gap closes at
$t/U\approx 0.26$, indicating a transition to the semi-metal phase.

To summarize, study of both magnetic and charge properties using series
expansions show a direct transition from the antiferromagnetic to the
semi-metal phase around $U_c\approx 4t$.

Combining the quantum monte carlo and series expansion results, we
estimate the phase transition to be in the range $U_c/t=4-5$. There
is greater internal consistency in the location of the
critical point if we restrict ourselves
to one method. But, in fact, there are larger uncertainties in both
methods especially as the quantum phase transition is reached. However,
both methods strongly indicate that the Mott transition and the antiferromagnetic
order happen simultaneously.

\section{Signatures of the Quantum Phase Transition in the Specific Heat}

An important objective of our study was to examine the signature of
the quantum phase transition in the finite temperature behavior of
the specific heat. We now turn to those studies, which are based
on the quantum monte carlo method.

At strong couplings, one expects two features
in the specific heat of the Hubbard Hamiltonian.  The first, at
a temperature $T \approx U/5$, signals the formation of
magnetic moments,\cite{Paiva01,Georges96} while the second, at a lower
temperature $T \approx J = 4t^2/U$, is associated with the
entropy of moment ordering.  This picture has been verified in
the one-dimensional case using Bethe Ansatz techniques\cite{Takahashi74}
and (using quantum monte carlo)
in the two dimensional square\cite{Duffy97,Staudt00}
and three dimensional cubic
lattices.\cite{Scalettar89}
Interestingly, in the square lattice, the two peak structure
persists to weak coupling where the energy scales
$U$ and $J$ have merged.\cite{Paiva01}
In one dimension, there is a single peak at weak
coupling.\cite{Shiba72,Schulte96}



The specific heat $C(T)$ for the two-dimensional honeycomb lattice is
shown in Fig.~16 for different couplings $U$ and lattice size $L=12$.
For strong coupling, $U/t=6,7,8$ there is a clear two peak structure.
This is replaced by a single peak for weaker couplings, $U/t=2,4,5$.
Again, this result is in contrast with the behavior of $C(T)$ on the square
lattice, where a two-peak structure is evident for all $U/T$.\cite{Paiva01}
It is plausible to conjecture that the difference is the absence
of long range antiferromagnetic order.
This suggestion is supported by the fact that
coalescence of the specific heat peaks is
seen in ``Dynamical Mean Field Theory''
(DMFT) \cite{Georges93,Vollhardt97,Chandra99} studies
when they are restricted to the paramagnetic phase and antiferromagnetic
fluctuations are neglected.

This is, however, a rather subtle question, since
the Mermin--Wagner theorem precludes long range order at
finite temperature.  What is meant, more precisely, is that
on a two-dimensional square lattice, the low $T$ structure in
$C(T)$ appears when
the antiferromagnetic correlation length
$\xi(T)$, begins to grow exponentially as $T\rightarrow 0$.

The evolution from a two to a one peak structure in $C(T)$ is
one interesting reflection of the underlying quantum
phase transition on the finite temperature thermodynamics.
Another way of examining this question concerns the low
temperature behavior of $C(T)$.
As pointed out in the introduction,
we expect a quadratic temperature dependence at both
strong coupling (spin-waves) and weak coupling (linearly vanishing
density of states at the Fermi level).
How does the coefficient $\delta$
in $C(T) = \delta T^2$ evolve as one crosses between the two phases?

\begin{figure}
\label{CvsTallU}
\includegraphics[width=2.8in,height=2.8in,angle=0]{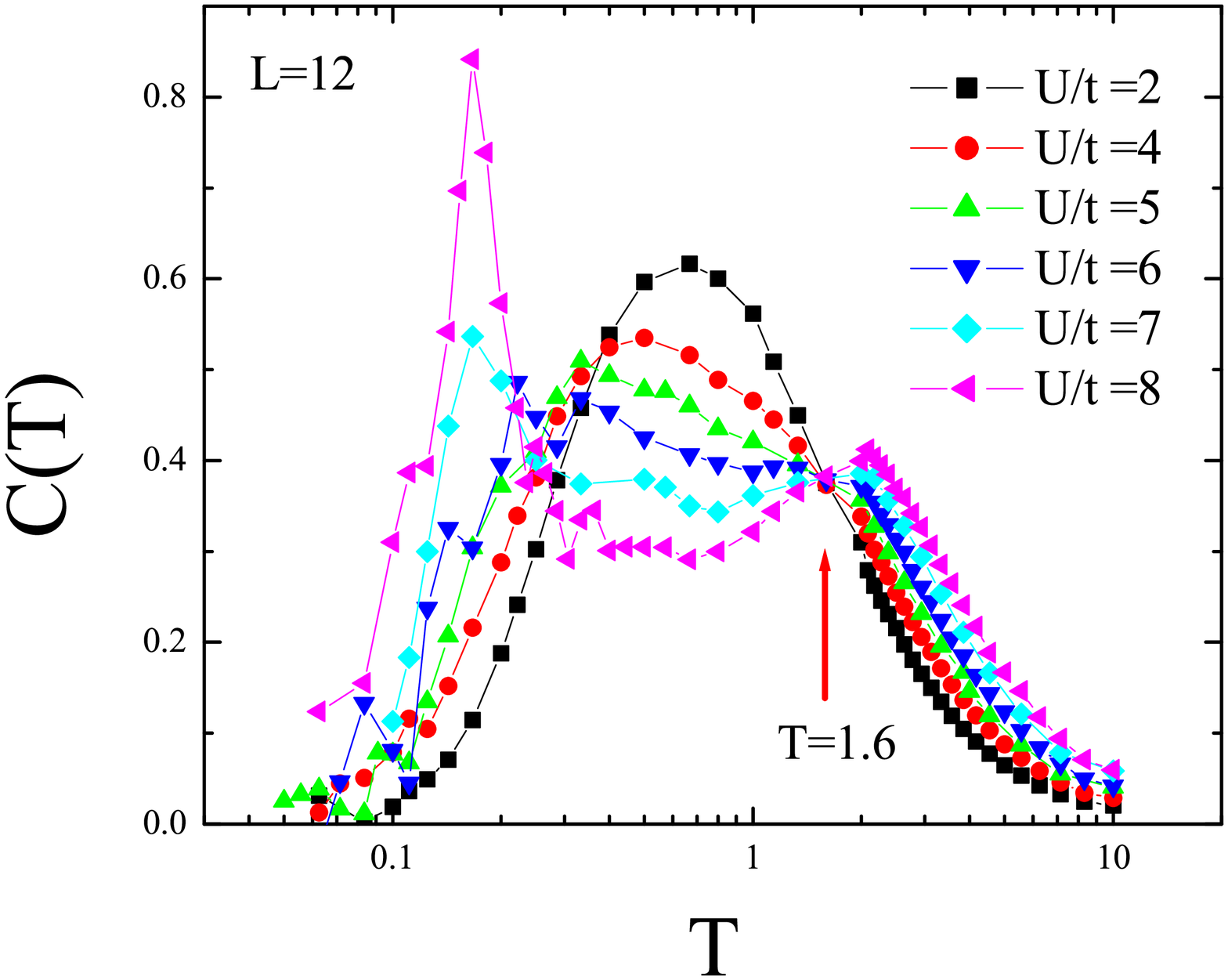}
\caption{
The specific heat $C(T)$ is shown as a function of temperature
for different coupling strengths.  In the antiferromagnetic
phase for $U > U_c$, the specific
heat has a two peak structure.  In the metallic phase for $U < U_c$
there is a single peak.  The `universal crossing' at $T=1.6t$ is
discussed in the text.
}
\end{figure}

Before we present the results for $\delta$, we note that
extracting $\delta$ is clearly a subtle numerical issue.  On the
one hand, $\delta$ characterizes the low $T$ behavior, but on the
other hand, because of finite size effects, which become larger as
the temperature is lowered, one cannot use data at too low values of $T$.
Thus, our calculation of $\delta$ should be viewed with some caution.
What we have done in generating Figs.~15-16 is to
fit the data for $C(T)$ to the $T^2$ form over only
a finite temperature window:  below the peak in $C(T)$
but also above the temperatures at which finite size effects
begin introducing a noticeable gap in the spectrum.
In Fig.~15 we show $\delta$ as a function of $U$.  There is a
structure in this plot in the vicinity of the value $U_c/t \approx 5$
previously inferred from the compressibility and spin correlation data.
Fig.~16 emphasizes this feature by plotting the derivative of $\delta$
with respect to $t/U$ as a function of $t/U$.
As we have noted, the specific heat of the noninteracting system
obeys $C=\delta(U=0)T^2$ with $\delta(U=0)=4.1$,
because of the linearly vanishing density of states.
Perturbation theory suggests that for small finite $U$,
$\delta$ should increase quadratically from this value.
Nevertheless, in the vicinity below the quantum phase transition,
the value for $\delta$ extracted from the
quantum monte carlo data looks rather linear in $U$, as seen in Fig. 15.
If $\delta = m U/t$ then $d(\delta)/d(t/U) = - m /(t/U)^2$.
With this in mind, a line showing the functional form
$-m / (t/U)^2$ with $m=2$ is given and
fits the weak coupling data
very well.  The breakaway from this form
at strong coupling further emphasizes the
change in behavior in the vicinity of the
quantum phase transition.

In studies of the two peak structure of the specific heat on the
square lattice, an interesting interchange of the role of
kinetic and potential energies was noted.\cite{Paiva01}
At large $U$, the temperature derivative of the potential energy
was the primary contribution to the high $T$, `moment formation', peak,
while the temperature derivative of the kinetic energy drove
the low $T$, `moment ordering', peak.  However, at weak $U$ the
situation was reversed, with the high $T$ peak originating in the
kinetic energy.  With that separation in mind, we plot in Fig.~17,
for the honeycomb lattice,
the contributions of the potential and kinetic energies  to
$\delta$.  It is the contribution
of the potential energy to $\delta$ which
appears to have the sharper evolution in the
vicinity of the
quantum phase transition.

Returning to the specific heat versus temperature, shown in Fig.~14,
we note the existence of a very well defined crossing point at
$T \approx 1.6t$.
This crossing has been observed previously in
DMFT,\cite{Georges93,Vollhardt97,Chandra99}
and in the two dimensional square lattice.\cite{Duffy97,Paiva01}
Indeed, in the former case, two crossings were observed, with the
high temperature one being nearly universal, while the low
temperature intersections were considerably more spread out,
much as we observe in Fig.~14.
It is also interesting that the numerical value of the crossing is
almost identical for the honeycomb
and square lattices, despite their different bandwidths.

\begin{figure}
\label{deltavsU}
\includegraphics[width=2.8in,height=2.8in,angle=0]{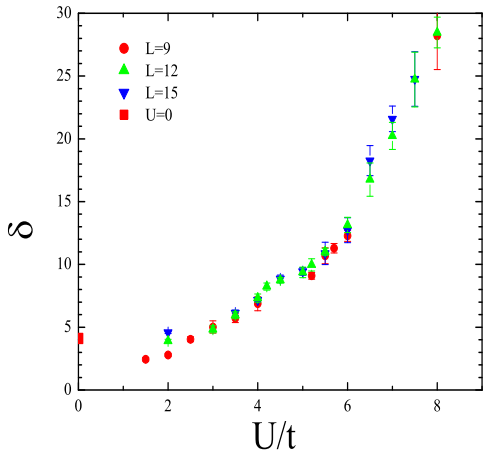}
\caption{
$\delta$, the coefficient of the $T^2$ term
in the specific heat is shown as a function of $U/t$.
The solid square is the $U=0$ value.
There appears to be a change
in slope as $U$ crosses $U_c$.
}
\end{figure}

\begin{figure}
\label{ddeltadtu}
\includegraphics[width=2.8in,height=2.8in,angle=0]{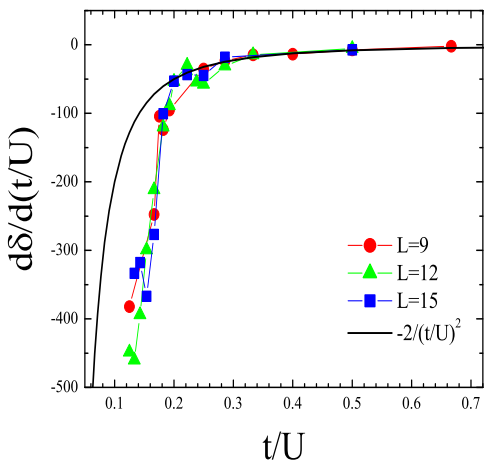}
\caption{
The derivative of $\delta$, the coefficient of the $T^2$ term
in the specific heat, with respect to $t/U$ is shown.  This
derivative has a sharp change near the critical coupling $U_c$.
The solid line is $-2/(t/U)^2$ (see text).
}
\end{figure}

\begin{figure}
\label{KandP}
\includegraphics[width=2.8in,height=2.8in,angle=0]{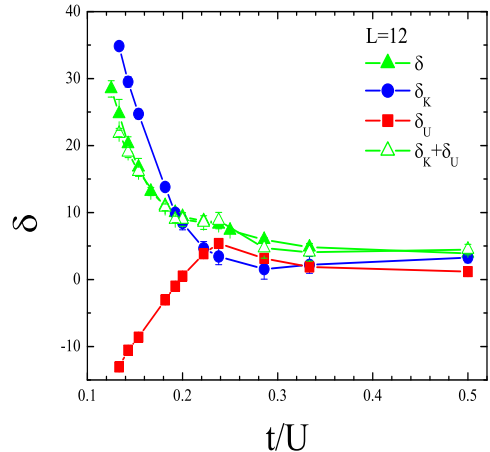}
\caption{
The separate contributions of the potential ($\delta_U$) and kinetic
($\delta_K)$ energies
to the quadratic coefficient of the specific heat are shown.
$\delta_U$ shows the more abrupt behavior in the vicinity of $U_c$.
The small differences between the values of $\delta$ obtained from
the total energy, and the values $\delta_K+\delta_U$ from the
kinetic and potential energies separately provide a measure of
the uncertainties
in our fitting procedure.
}
\end{figure}

Finally, we turn to the behavior of the entropy $S$.  In Fig.~18 we
show $S$ as a function of $U$ for different temperatures $T$.
At large $U$, the clustering of the curves for different temperatures
near ln(2) is indicative of the existence of
disordered magnetic moments in a range of intermediate $T$.
The low temperature magnetic ordering tendency is evident in the
gap between the $T=0.2$ and $T=0.3$ curves.
As $U$ is decreased, the screening away of the
moments is indicated by the $T=0.3$ isotherm dropping from ln(2) to 0.
It is interesting that this behavior is so gradual.
Finally at small $U$ one observes the more or less equally spaced isotherms of
free electron gas.  This figure
complements the data of $C(T)$ shown in Fig.~14, since the entropy hang
up at large U near ln(2) is just the $C/T$ area of the lower specific
heat peak.

Figure 19 exhibits the entropy as a function of temperature.
At weak coupling, there is a smooth evolution from ln(4) at
high $T$ to zero at low $T$.  For strong coupling, a plateau
near ln(2) interrupts this evolution, again exhibiting a range
of temperatures with well formed, but disordered moments.

\begin{figure}
\label{SvsU}
\includegraphics[width=2.8in,height=2.8in,angle=0]{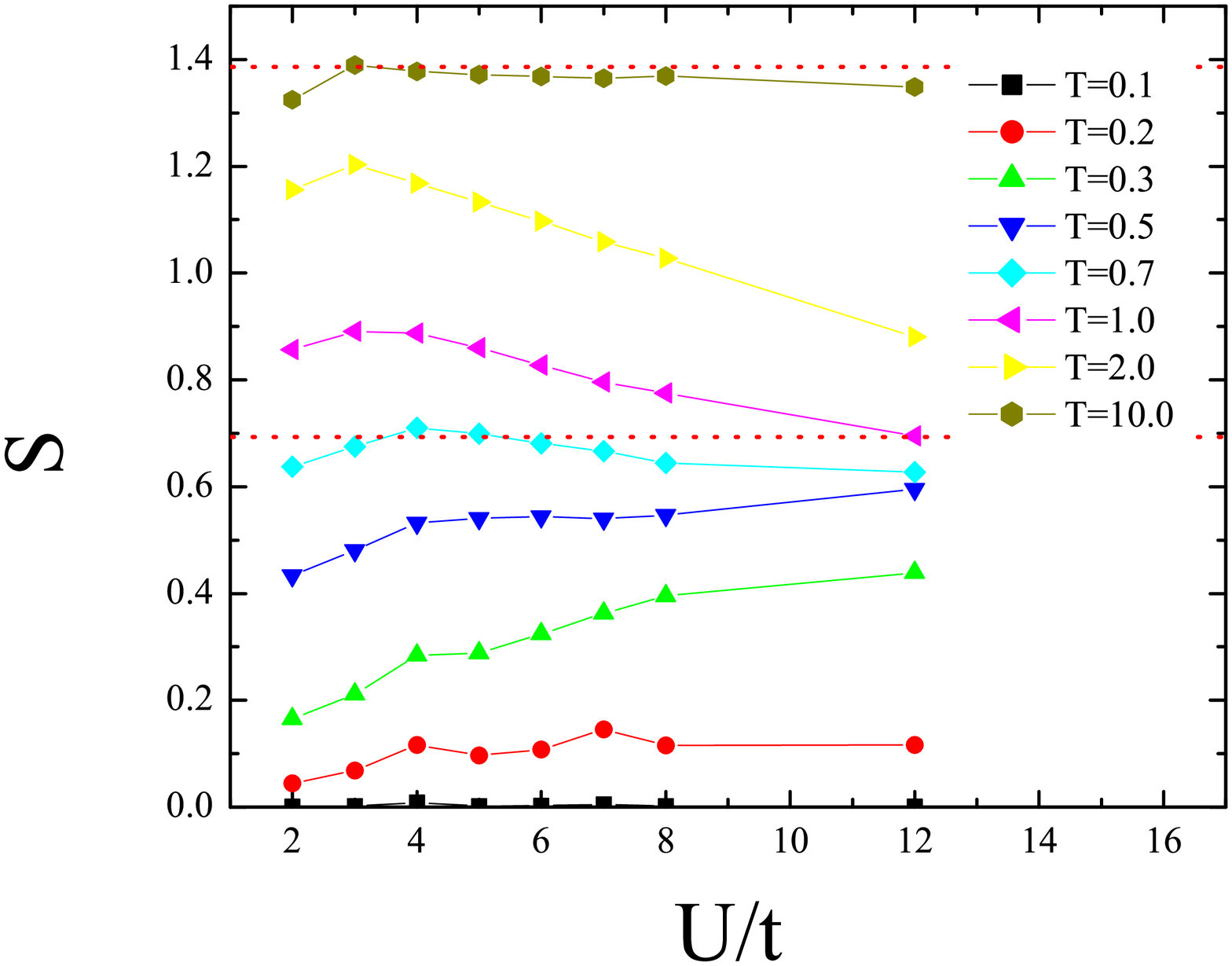}
\caption{
The entropy is shown as a function of $U$ for different temperatures.
At large $U$ the gaps between the $T=10$ and $T=2$ curves and between the
$T=0.3$ and $T=0.2$ curves reflect the entropy loss associated with
magnetic moment formation and ordering respectively.
}
\end{figure}

\begin{figure}
\label{entropy2.EPS}
\includegraphics[width=2.8in,height=2.8in,angle=0]{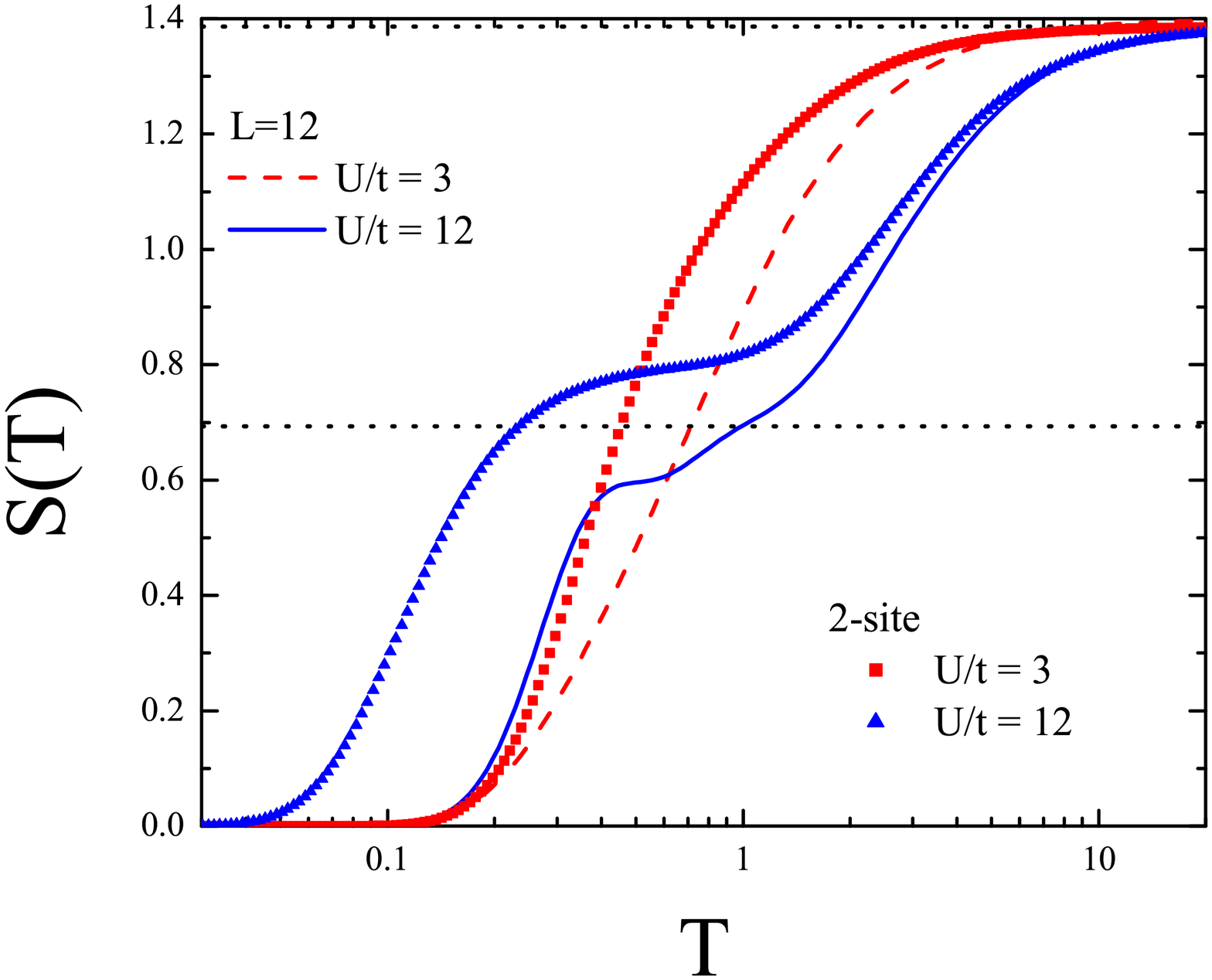}
\caption{
The entropy is shown as a function of $T$ for weak and strong coupling.
The dashed and solid lines are the results of quantum monte carlo simulations
on 12x12 lattices.  The symbols are generated by an exact calculation
on a two site model for comparison.
}
\end{figure}

\section{Conclusions}

In this paper we have studied the Hubbard Hamiltonian on
a half--filled honeycomb lattice using quantum monte carlo
and series expansion methods. Both methods strongly suggest
that the model has a single continuous transition at $T=0$,
between an antiferromagnetic phase at large $U/t$ and a
semi-metal phase at small $U/t$. Quantum monte carlo
results for the compressibility, which looks at the charge response
of the system, and the magnetic structure factor, which looks at
the spin response, both suggest a transition
around $U_c/t\approx 5$. The series expansion results
for the sublattice magnetization, which is the spin order
parameter and the charge excitation gap, which characterizes
the Mott transition, both point to a single transition at $U/t\approx 4$.
The discrepency between the quantum monte carlo and
series expansion results reflects the uncertainties in the
calculations, especially as the critical point is approached.
Thus we expect the transition to lie in the range $4<U/t<5$,
a result in complete agreement with the previous work of Martelo
et al. \cite{Martelo97}.

Finally, one of the goals of this work was to look
for finite temperature signatures of the phase transition
in the specific heat, as a guide to experimental studies. We observe that
around $U_c$ the specific heat changes from a one peak (below
$U_c$) to a two peak (above $U_c$) structure.  We suggest that this is
associated with the fact that for $U>U_c$ the antiferromagnetic
correlation length grows rapidly as the temperature is reduced.
For weak coupling only very short-range antiferromagnetic correlations exist,
and the specific heat has no signature of magnetic order.

We also studied the evolution with on-site interaction
strength $U$ of the coefficient $\delta(U)$ of the quadratic
temperature dependence of the specific heat at low temperatures.
Since the excitations which produce the $T^2$ term above and below
the quantum phase transition are unrelated,
one might have expected $\delta(U)$ to exhibit a discontinuity at $U_c$.
Instead, we found a sharp change in the slope, $d \delta(U) / dU$
at $U_c$.  Given the uncertainties in obtaining $\delta(U)$, from finite-size
calculations, these results should be viewed with some caution.
Experimental searches for such a behavior would be quite interesting.

\vskip0.2in
\noindent
\acknowledgments
We acknowledge very useful conversations with W.E.~Pickett
and A.K.~McMahan.
This work was supported by the CNPq--Brazil and FUJB--Brazil (TP),
US National Science Foundation grants DMR--0312261 (RTS),
INT--0203837 (RTS), and DMR--0240918 (RRPS), and by
a grant from the Australian Research Council (WZ and JO).
W.Z.  wishes to thank the University of California at Davis  for hospitality
while part of the work was being done.
 We are grateful for the computing resources provided
 by  the Australian Centre for Advanced Computing and Communications (AC3)
and by the Australian Partnership for Advanced Computing (APAC)
National Facility.


\begin{widetext}

\begin{table}
\squeezetable
\caption{
Series coefficients for Ising expansions of the ground-state energy per site,
$E_0/NU$, the staggered magnetization $M$, squared local moment $L_m$, and
the uniform magnetic susceptibility
$\chi_{\perp}$ for $t/U=0.15$ and $J/U=0.0225$.
Coefficients of $\lambda^n$
up to order $\lambda^{15}$ are listed.
  } \label{tab1}
\begin{ruledtabular}
\begin{tabular}{|c|l|l|l|l|}
 \multicolumn{1}{|c|}{$n$} &\multicolumn{1}{c|}{$E_0/NU$}
 &\multicolumn{1}{c|}{$M$}
 &\multicolumn{1}{c|}{$L_m$}
 &\multicolumn{1}{c|}{$\chi_{\perp}$}\\
\hline
  0 & -0.250000000     & ~1.000000000     & ~1.000000000     & ~3.703703704     \\
  1 & ~0.000000000     & ~0.000000000     & ~0.000000000     & ~3.703703704     \\
  2 & -6.067415730$\times 10^{-2}$ & -1.090771367$\times 10^{-1}$ & -1.090771367$\times 10^{-1}$   & -4.743810362     \\
  3 & -6.135588941$\times 10^{-3}$~~~ & -2.206054451$\times 10^{-2}$~~~ & -2.206054451$\times 10^{-2}$~~~ & -1.394196294$\times 10^{1}$~~~ \\
  4 & -1.172455577$\times 10^{-2}$ & -3.552865175$\times 10^{-1}$ & -3.819372077$\times 10^{-2}$ & -5.304439204     \\
  5 & -1.595132209$\times 10^{-2}$ & -7.348348039$\times 10^{-1}$ & -6.041765980$\times 10^{-2}$ & ~2.402576623$\times 10^{1}$ \\
  6 & -2.167350013$\times 10^{-3}$ & -4.545288062$\times 10^{-1}$ & ~1.617075260$\times 10^{-2}$ & ~3.836014584$\times 10^{1}$ \\
  7 & ~1.484188843$\times 10^{-2}$ & ~5.706090877$\times 10^{-1}$ & ~1.182931503$\times 10^{-1}$ & -6.261816470     \\
  8 & ~1.982803975$\times 10^{-2}$ & ~1.508870723     & ~1.282196638$\times 10^{-1}$ & -8.556886959$\times 10^{1}$ \\
  9 & ~1.098873088$\times 10^{-2}$ & ~1.362697325     & ~2.870390056$\times 10^{-2}$ & -8.473956953$\times 10^{1}$ \\
 10 & -5.309942069$\times 10^{-3}$ & -9.038107499$\times 10^{-2}$ & -1.194968503$\times 10^{-1}$ & ~7.394366538$\times 10^{1}$ \\
 11 & -2.087445585$\times 10^{-2}$ & -1.988476084     & -2.381637820$\times 10^{-1}$ & ~2.479347440$\times 10^{2}$ \\
 12 & -2.733805940$\times 10^{-2}$ & -2.967041815     & -2.434002720$\times 10^{-1}$ & ~1.283247884$\times 10^{2}$ \\
 13 & -1.868451700$\times 10^{-2}$ & -2.220001906     & -7.558507985$\times 10^{-2}$ & -3.404916769$\times 10^{2}$ \\
 14 & ~4.985233235$\times 10^{-3}$ & -1.210494842$\times 10^{-2}$ & ~2.437663805$\times 10^{-1}$ &  \\
 15 & ~3.577717123$\times 10^{-2}$ & ~2.934366446     & ~5.886265926$\times 10^{-1}$ &  \\
\end{tabular}
\end{ruledtabular}
\end{table}

\begin{table}
\caption{
Series coefficients for the  spin-wave excitation spectrum $ \Delta (k_x ,k_y)/U$ and
1-hole dispersion $ \Delta_{\rm 1h} (k_x ,k_y)/U$. Nonzero coefficients
up to order $\lambda^{13}$ for $t/U=0.15$ and $J/U=0.0225$ are listed.
  } \label{tab2}
\begin{ruledtabular}
\begin{tabular}{|cl|cl|cl|cl|}
 \multicolumn{1}{|c}{$(i,j,p)$} &\multicolumn{1}{c}{$a_{i,j,p}$} &
 \multicolumn{1}{|c}{$(i,j,p)$} &\multicolumn{1}{c}{$a_{i,j,p}$} &
 \multicolumn{1}{|c}{$(i,j,p)$} &\multicolumn{1}{c}{$a_{i,j,p}$} &
 \multicolumn{1}{|c}{$(i,j,p)$} &\multicolumn{1}{c|}{$a_{i,j,p}$} \\
\hline
\multicolumn{8}{|c|}{spin-wave excitation spectrum $ \Delta (k_x ,k_y)/U$ } \\
 ( 0, 0, 0) & ~1.350000000$\times 10^{-1}$ &(10, 0, 0) & ~2.216091139$\times 10^{-1}$ &(10, 3, 1) & ~4.746162497$\times 10^{-1}$ &( 8, 6, 0) & -7.225113429$\times 10^{-3}$ \\
 ( 1, 0, 0) & -1.350000000$\times 10^{-1}$ &(11, 0, 0) & -1.393969235$\times 10^{-2}$ &(11, 3, 1) & -1.099204748$\times 10^{-1}$ &( 9, 6, 0) & -2.262498583$\times 10^{-2}$ \\
 ( 2, 0, 0) & ~1.111175434$\times 10^{-1}$ &(12, 0, 0) & -5.637616584$\times 10^{-1}$ &(12, 3, 1) & -1.440323067   &(10, 6, 0) & -2.696906952$\times 10^{-2}$ \\
 ( 3, 0, 0) & ~2.082046617$\times 10^{-2}$ &(13, 0, 0) & -1.304334460   &(13, 3, 1) & -3.353557841   &(11, 6, 0) & ~6.591260743$\times 10^{-3}$ \\
 ( 4, 0, 0) & -4.424242442$\times 10^{-2}$ &( 4, 3, 1) & -1.237981890$\times 10^{-1}$ &( 8, 6, 2) & -3.350958277$\times 10^{-3}$ &(12, 6, 0) & -2.916010244$\times 10^{-3}$ \\
 ( 5, 0, 0) & -3.676674665$\times 10^{-2}$ &( 5, 3, 1) & -1.341231436$\times 10^{-1}$ &( 9, 6, 2) & -1.098231382$\times 10^{-2}$ &(13, 6, 0) & -2.568183304$\times 10^{-1}$ \\
 ( 6, 0, 0) & -9.749111612$\times 10^{-3}$ &( 6, 3, 1) & -2.171649093$\times 10^{-2}$ &(10, 6, 2) & -1.153863507$\times 10^{-2}$ &(12, 9, 3) & -1.609104960$\times 10^{-3}$ \\
 ( 7, 0, 0) & ~2.153747460$\times 10^{-2}$ &( 7, 3, 1) & ~1.086787464$\times 10^{-1}$ &(11, 6, 2) & ~7.460180712$\times 10^{-3}$ &(13, 9, 3) & -8.451196579$\times 10^{-3}$ \\
 ( 8, 0, 0) & ~9.813204715$\times 10^{-2}$ &( 8, 3, 1) & ~2.930587649$\times 10^{-1}$ &(12, 6, 2) & ~7.410248941$\times 10^{-3}$ &(12, 9, 1) & -9.742145152$\times 10^{-3}$ \\
 ( 9, 0, 0) & ~2.080499557$\times 10^{-1}$ &( 9, 3, 1) & ~5.034160236$\times 10^{-1}$ &(13, 6, 2) & -1.092694552$\times 10^{-1}$ &(13, 9, 1) & -5.101138671$\times 10^{-2}$ \\
\hline
\multicolumn{8}{|c|}{1-hole dispersion $ \Delta_{\rm 1h} (k_x ,k_y)/U$ } \\
 ( 0, 0, 0) & ~5.675000000$\times 10^{-1}$ &( 7, 3, 1) & ~5.141564645   &( 8, 6, 0) & -1.091561629   &( 9,12, 4) & -2.562349516$\times 10^{-6}$ \\
 ( 1, 0, 0) & -6.750000000$\times 10^{-2}$ &( 8, 3, 1) & -1.258196166$\times 10^{1}$ &( 9, 6, 0) & -5.455143302   &(10,12, 4) & ~2.676464885$\times 10^{-3}$ \\
 ( 2, 0, 0) & -6.336614782$\times 10^{-1}$ &( 9, 3, 1) & -1.733463014$\times 10^{2}$ &(10, 6, 0) & ~5.327561412   &(11,12, 4) & ~1.433265640$\times 10^{-2}$ \\
 ( 3, 0, 0) & -7.336392902$\times 10^{-1}$ &(10, 3, 1) & -6.526403290$\times 10^{1}$ &(11, 6, 0) & ~8.443452858$\times 10^{1}$ &( 8,12, 2) & -2.978280918$\times 10^{-5}$ \\
 ( 4, 0, 0) & ~2.938067768   &(11, 3, 1) & ~3.901815114$\times 10^{3}$ &( 6, 9, 3) & ~8.189533379$\times 10^{-5}$ &( 9,12, 2) & -2.049879613$\times 10^{-5}$ \\
 ( 5, 0, 0) & ~1.013199055$\times 10^{1}$ &( 4, 6, 2) & -2.251112570$\times 10^{-3}$ &( 7, 9, 3) & ~3.983472192$\times 10^{-5}$ &(10,12, 2) & ~2.139465782$\times 10^{-2}$ \\
 ( 6, 0, 0) & -1.702110647$\times 10^{1}$ &( 5, 6, 2) & -6.420330616$\times 10^{-4}$ &( 8, 9, 3) & ~1.025604453$\times 10^{-2}$ &(11,12, 2) & ~1.146460686$\times 10^{-1}$ \\
 ( 7, 0, 0) & -1.514268463$\times 10^{2}$ &( 6, 6, 2) & ~2.127063569$\times 10^{-2}$ &( 9, 9, 3) & ~4.129584046$\times 10^{-2}$ &( 8,12, 0) & -2.233710688$\times 10^{-5}$ \\
 ( 8, 0, 0) & -4.655833834   &( 7, 6, 2) & ~6.346958017$\times 10^{-2}$ &(10, 9, 3) & -2.110091800$\times 10^{-1}$ &( 9,12, 0) & -1.537409710$\times 10^{-5}$ \\
 ( 9, 0, 0) & ~2.217144323$\times 10^{3}$ &( 8, 6, 2) & -5.148299650$\times 10^{-1}$ &(11, 9, 3) & -1.677872051   &(10,12, 0) & ~1.604362363$\times 10^{-2}$ \\
 (10, 0, 0) & ~3.740789745$\times 10^{3}$ &( 9, 6, 2) & -2.603465931   &( 6, 9, 1) & ~4.913720027$\times 10^{-4}$ &(11,12, 0) & ~8.598244287$\times 10^{-2}$ \\
 (11, 0, 0) & -2.947388888$\times 10^{4}$ &(10, 6, 2) & ~2.034643910   &( 7, 9, 1) & ~2.390083315$\times 10^{-4}$ &(10,15, 5) & ~1.895585905$\times 10^{-7}$ \\
 ( 2, 3, 1) & ~1.238532110$\times 10^{-1}$ &(11, 6, 2) & ~3.720022590$\times 10^{1}$ &( 8, 9, 1) & ~6.168519384$\times 10^{-2}$ &(11,15, 5) & ~1.687061931$\times 10^{-7}$ \\
 ( 3, 3, 1) & ~1.022641192$\times 10^{-2}$ &( 4, 6, 0) & -4.502225139$\times 10^{-3}$ &( 9, 9, 1) & ~2.478775229$\times 10^{-1}$ &(10,15, 3) & ~1.895585905$\times 10^{-6}$ \\
 ( 4, 3, 1) & ~7.872629131$\times 10^{-2}$ &( 5, 6, 0) & -1.284066123$\times 10^{-3}$ &(10, 9, 1) & -1.373111288   &(11,15, 3) & ~1.687061931$\times 10^{-6}$ \\
 ( 5, 3, 1) & ~8.219599490$\times 10^{-3}$ &( 6, 6, 0) & ~4.204959522$\times 10^{-2}$ &(11, 9, 1) & -1.064077735$\times 10^{1}$ &(10,15, 1) & ~3.791171810$\times 10^{-6}$ \\
 ( 6, 3, 1) & ~6.643492076$\times 10^{-1}$ &( 7, 6, 0) & ~1.267005690$\times 10^{-1}$ &( 8,12, 4) & -3.722851147$\times 10^{-6}$ &(11,15, 1) & ~3.374123863$\times 10^{-6}$ \\
\end{tabular}
\end{ruledtabular}
\end{table}
\end{widetext}

\end{document}